\documentclass{Interspeech2024}

\interspeechcameraready 
\usepackage{tikz}
\usepackage{collcell}
\usepackage{multirow}
\usepackage{amsmath}
\title{Emo-bias: A Large Scale Evaluation of Social Bias on Speech Emotion Recognition}

\name[affiliation={1}]{Yi-Cheng}{Lin}
\name[affiliation={1}]{Haibin}{Wu}
\name[affiliation={2}]{Huang-Cheng}{Chou}
\name[affiliation={2}]{Chi-Chun}{Lee}
\name[affiliation={1}]{Hung-yi}{Lee}

\address{
  $^1$National Taiwan University, Taiwan\\
  $^2$National Tsing Hua University, Taiwan
}
\email{\{r12942075,hungyilee\}@ntu.edu.tw,
cclee@ee.nthu.edu.tw}


\begin{document}

\maketitle

\begin{abstract}
The rapid growth of Speech Emotion Recognition (SER) has diverse global applications, from improving human-computer interactions to aiding mental health diagnostics.
However, SER models might contain social bias toward gender, leading to unfair outcomes. 
This study analyzes gender bias in SER models trained with Self-Supervised Learning (SSL) at scale, exploring factors influencing it. SSL-based SER models are chosen for their cutting-edge performance.
Our research pioneering research gender bias in SER from both upstream model and data perspectives.
Our findings reveal that females exhibit slightly higher overall SER performance than males. 
Modified CPC and XLS-R, two well-known SSL models, notably exhibit significant bias.
Moreover, models trained with Mandarin datasets display a pronounced bias toward valence.
Lastly, we find that gender-wise emotion distribution differences in training data significantly affect gender bias, while upstream model representation has a limited impact. 
\end{abstract}

\section{Introduction}
Speech emotion recognition (SER) aims to detect and interpret emotional states conveyed through speech signals. 
However, SER models may capture and learn social bias, leading to potential social harm. 
Biased SER systems may exacerbate existing inequalities by disproportionately affecting marginalized communities. 
For example, if a system is less accurate in recognizing emotions in individuals with disabilities, non-native speakers, or a specific gender, it could further marginalize these groups by denying them equitable access to services or opportunities.

While extensive research has addressed bias in various machine learning domains, such as Automatic Speech Recognition (ASR) \cite{ngueajio2022hey, feng2021quantifying, attanasio2024multilingual}, Speech Translation \cite{savoldi2022under, gaido2021split}, Facial Emotion Recognition \cite{domnich2021responsible}, and Automatic Speaker Verification (ASV) \cite{fenu2020exploring}, limited attention has been paid to social bias within SER systems. 
For instance, Gorrostieta et al. \cite{gorrostieta2019gender} evaluated gender bias within a specific model and dataset, proposing two adversarial debiasing approaches. 
However, their analysis was confined to a singular model and only one dataset, potentially limiting its applicability to broader contexts.
Similarly, Chien et al. \cite{10094984} investigated gender-specific emotion perception using the IEMOCAP dataset \cite{Busso_2008_5}, presenting a perceptual emotion learning framework. 
Yet, they overlooked the impact of training dataset selection on emotional bias. 
This underscores the need for comprehensive investigations into gender bias across diverse SER models and datasets to ensure robust and generalizable results.

Recognizing such research gaps, our study delves into two inquiries: 
Firstly, do contemporary SER models exhibit gender bias? 
Secondly, what are the primary factors contributing to such bias? 
Specifically, we investigate whether upstream representations and downstream training data play a crucial role in shaping bias within these models.

Leveraging the cutting-edge advancements in speech self-supervised learning (SSL) \cite{mohamed2022self, atmaja2022evaluating}, we employ 15 SSL models and classical speech features like FBank to train SER systems. 
Through rigorous and comprehensive experimentation across six diverse emotion datasets, we carefully train and assess a total of 96 SER models. 

Our work yields the following contributions:
\begin{itemize}
    \item We conduct a large-scale evaluation of 15 SER models on six emotion datasets. Notably, females exhibit slightly superior overall emotion recognition performance to males, with a substantial gender-wise performance gap evident across individual emotions. Our analysis highlights two models, Modified CPC \cite{riviere2020unsupervised} and XLS-R \cite{babu2021xls}, as exhibiting the highest bias in gender-wise SER $F_1$-score differences. 
    \item  We found that downstream training data distribution significantly affects gender bias for models trained with acted datasets while having a medium correlation with SER performance bias for real-world datasets.
    \item We analyze the gender-wise $F_1$-score difference on valence. We observe that ``females have higher $F_1$-score on positive valence while males have higher $F_1$-score on negative valence" is apparent in models trained with Chinese datasets.
    \item We analyze the correlation between \textit{gender-wise upstream representation bias} and \textit{SER performance gap between two genders}. We find that SSL upstream representations barely influence the bias in SER performance.
\end{itemize} 

\section{Evaluation design}


\subsection{SSL model and downstream model}
We chose SSL-based SER systems to evaluate gender bias, as these SOTA models are commonly preferred for emotion evaluation \cite{10089511, 9747870}, potentially amplifying the impact of any bias present.
In the framework of speech SSL, model training consists of two stages. 
The first stage pre-trains a model (or upstream model) using self-supervised learning with a predefined pretext objective. 
The second stage uses representation from the upstream model to train a downstream task such as SER.

We use the SSL-based SER models in \cite{wu2024emosuperb}, which uses publicly available speech SSL models collected by the S3PRL toolkit \cite{yang21c_interspeech}.
The upstream models include models trained with generative approach: DeCoAR 2 \cite{Ling_2020}, Autoregressive Predictive Coding (APC) \cite{Chung_2019}, Vector-Quantized APC (VQ-APC) \cite{Chung_2020}, Nonautoregressive Predictive Coding (NPC) \cite{liu2020non}, TERA \cite{Liu_2021}, Mockingjay (Mock) \cite{Liu_2020}; models trained with contrastive approach: Wav2Vec2-XLS-R-1B (XLS-R) \cite{babu2021xls}, Wav2Vec 2.0 (W2V2) \cite{baevski2020wav2vec}, Wav2Vec2 Large Robust (W2V2 R) \cite{hsu21_interspeech}, vq-wav2vec (VQ-W2V) \cite{Baevski_2020_2}, wav2vec (W2V) \cite{schneider2019wav2vec}, and Contrastive Predictive Coding (M CPC) \cite{riviere2020unsupervised}; models trained with predictive approach: HuBERT \cite{hsu2021hubert}, WavLM \cite{Chen_2022}, Data2Vec (D2V) \cite{pmlr-v162-baevski22a}. 

We use the same downstream model architecture as the SER task in the S3PRL toolkit \cite{yang21c_interspeech}, using three Conv1d, a self-attention pooling, and two linear layers.
To capture the high-dimensional nature of emotion \cite{Cowen_2021}, we formulate emotion recognition as a multilabel classification problem. 
We first transform the emotion annotations to emotion distribution by frequency and then apply label smoothing by \cite{Szegedy_2016} using a smoothing parameter of 0.05 to obtain a soft label. 
We use the $F_1$ score as the evaluation metric for performance, which aligns with SER challenges
 \cite{Goncalves_2022} and benchmarks \cite{wu2024emosuperb}. 
A label prediction is successful if the output emotion probability distribution is higher than $\frac{1}{n}$ for $n$ emotion classification task. Our models are trained on Nvidia Tesla V100 GPUs with 32 GB of memory. The cumulative GPU runtime amounts to approximately 3,300 hours.

\subsection{Bias evaluation}
In our study, we employ three evaluation metrics to assess gender bias. 
We hope our evaluation plan can serve as a valuable reference for others.

\subsubsection{$F_1$ score difference}
We quantify the gender bias $d_{e_i}$ on each emotion $e_i$ by the difference of $F_1$ score on gender because $F_1$ score reflects both precision and recall. Furthermore, we define the corpus-level gender bias $d_c$ by averaging the absolute value of $d_{e_i}$ for all emotions $e_i$ in the set of all emotion $E$ in the corpus:
\begin{align}
    d_{e_i}& = F_{1,e_i female}-F_{1 ,e_i male}\\
    d_{c}& = mean_{e_i\in E} |d_{e_i}|.
\end{align}
We further evaluate the gender bias on valence by Eq.~\ref{d_v}.\footnote{We don't use absolute value on $d_v$ as $d_c$, because we mean to compare it with upstream bias in section~\ref{subsubsection: upstream}} The intuition is calculating the difference of $d_{e_i}$ from positive valence and negative valence:
\begin{align}
    d_{v}& = \Sigma_{e_i\in E_+}d_{e_i}-\Sigma_{e_i\in E_-}d_{e_i} \notag \\ 
    &+ \Sigma_{e_i\in E_b} (p_{+}d_{e_i,+} - p_{-}d_{e_i,-}).
    \label{d_v}
\end{align}
\begin{table}[t]
\centering
\fontsize{7}{9}\selectfont
\caption{Overview of mapping between valence and emotion. Surprise can have a positive or negative valence.}
\vspace{-1em}
\label{tab:emotion_valence}
\setlength{\tabcolsep}{2pt}
\renewcommand{\arraystretch}{0.9}
\begin{tabular}{c|c}
\toprule
\textbf{Valence} & \textbf{Emotion} \\
\midrule
positive & Happiness, Excitement, Relax, Joy \\
\midrule
negative & Anger, Disgust, Contempt, Frustration, Disappointment, Sadness, Fear\\
\midrule
both & Surprise\\
\bottomrule
\end{tabular}
\vspace{-0.3cm}
\end{table}

\hspace{-6pt}$E_+$, $E_-$, and $E_b$ are the set of emotions in the corpus that belong to positive valence, negative valence, and both valence, respectively. $p_+$ and $p_-$ are the portions of speech belonging to positive and negative valence. $d_{e_i,+}$ and $d_{e_i,-}$ are the $d_{e_i}$ of positive and negative valence speech, respectively. The detail of valence categorization is shown in Table~\ref{tab:emotion_valence}, following the study \cite{Cowen_2019}.

\subsubsection{Upstream representation bias on valence}
\label{subsubsection: upstream}
We would like to know whether gender bias in the upstream model embedding propagates to downstream applications. To evaluate how the upstream model's embedding of stimuli representing females and males relates to its embedding of stimuli representing positive and negative valence, we use the Speech Embedding Association Test (SpEAT) \cite{slaughter-etal-2023-pre} for detecting bias. SpEAT measures the relative cosine similarity between 4 groups of stimuli. Let X and Y represent sets of embeddings from female and male, and let A and B be sets of embeddings for positive and negative valence, respectively. SpEAT effect size $d$ is the difference between sums over the respective target concept, normalized by the standard deviation to compute the magnitude of association:
\begin{align}
    d_s& = \frac{\Sigma_{x\in X} s(x, A, B) - \Sigma_{y\in Y}s(y, A, B)}{std\_dev_{w\in X\cup Y} s(w, A, B)},
\end{align}
where each term is the difference between the mean of cosine similarity of gender to each valence:
\begin{align}
    s(w, A, B)& = mean_{a\in A} cos(w, a) - mean_{b\in B} cos(w, b).
\end{align}
In SpEAT, the embedding of speech segments is first done by averaging embedding in each layer and then averaging the aggregated embeddings across all layers, which we call the \textbf{Mean} aggregation. We further incorporate representation weights from SER models to achieve compatibility and comparability with the SSL paradigm. Assume SSL SER models with $n$ layers are trained to use $c_i$ as the weight for the input representation weighted sum of $i^{th}$ layer. We use $c_i$ to weighted sum embeddings across layers, which is called \textbf{Weighted} aggregation. This enhances the similarity of our representations to those generated by ER models trained via SSL. We average $c_i$ from SER models trained in different folds for cross-fold validation datasets defined in \cite{wu2024emosuperb}. 

\begin{table*}[t!]
\centering
\fontsize{7}{9}\selectfont
\caption{SER bias on emotion $d_e$ across 6 emotion datasets and 9 models, in \%. The emotions are abbreviated as follows. Angry: Ang, Disgusting: Disg, Contempt: Cont, Neutral: Neu, Surprise: Sur, Hap: Happy, Frustrated: Fru, Excited: Exc, Disappointed: Disa, Relax: Rel. \textbf{Mac} represents the macro-$F_1$ score over all emotions}
\vspace{-0.3cm}
\setlength{\tabcolsep}{2pt}
\renewcommand{\arraystretch}{0.83}
\begin{tabular}{l||ccccccccc|ccccccc|ccccccccccc}
\toprule
 & \multicolumn{9}{c}{BIIC-PODCAST} & \multicolumn{7}{|c|}{CREMA-D} & \multicolumn{10}{c}{IEMOCAP}\\
\midrule
 Model & Ang & Sad & Disg & Cont & Fear & Neu & Sur & Hap & \textbf{Mac} & Ang & Sad & Disg & Fear & Neu & Hap & \textbf{Mac} & Fru & Ang & Sad & Disg & Exc & Fear & Neu & Sur & Hap & \textbf{Mac}\\
\midrule
XLS-R & 6.3 & 12.0 & 0.0 & -5.3 & -0.9 & -24.7 & 2.3 & 27.4 & 2.1 & 2.2 & 11.7 & -2.8 & 2.9 & -2.6 & 12.6 & 4.0 & -0.4 & 7.2 & 2.2 & -1.7 & 1.8 & 8.1 & -2.5 & 6.5 & 7.7 & 3.2 \\
WavLM & 1.8 & 12.1 & 0.0 & 0.7 & 5.1 & -24.1 & 7.7 & 28.0 & 2.9 & 3.7 & 11.0 & 1.3 & 3.3 & -2.6 & 8.7 & 4.2 & -0.6 & 6.7 & 0.7 & 0.0 & 2.5 & 8.9 & -2.7 & 3.0 & 10.0 & 3.2 \\
W2V2 R & 6.6 & 7.7 & 0.0 & -5.9 & 0.0 & -25.2 & 6.7 & 28.9 & 2.3 & 3.1 & 11.8 & -3.0 & 3.0 & -2.6 & 13.4 & 4.3 & 0.4 & 6.3 & 2.0 & -0.6 & 1.5 & 3.1 & -3.5 & 3.8 & 7.9 & 2.3 \\
W2V2 & -3.3 & 9.5 & 0.0 & -7.8 & 1.9 & -27.6 & 8.3 & 28.8 & 1.2 & 1.2 & 6.4 & 0.1 & 4.6 & -2.7 & 5.7 & 2.6 & -1.2 & 5.3 & 2.0 & 0.0 & 0.5 & 1.0 & -7.0 & 0.0 & 7.5 & 0.9 \\
VQ-APC & -1.2 & 5.0 & 0.0 & -6.3 & 0.0 & -24.7 & 4.0 & 29.7 & 0.9 & -1.9 & 6.9 & -5.2 & 0.7 & -3.1 & 10.7 & 1.3 & -0.2 & 3.3 & 2.5 & 0.0 & 1.9 & 9.7 & -4.0 & -1.1 & 8.0 & 2.2\\
HuBERT & 3.8 & 13.4 & 0.0 & -6.3 & 2.7 & -26.1 & 4.4 & 29.3 & 2.6 & 2.0 & 8.0 & -1.5 & 1.9 & -2.3 & 7.9 & 2.7 & -0.3 & 6.4 & 2.5 & -7.6 & -2.0 & 2.7 & -3.0 & 3.9 & 7.5 & 1.1\\
DeCoAR 2 & 3.9 & 2.4 & 0.0 & -8.5 & 0.0 & -24.9 & 1.7 & 29.3 & 0.5 & 1.9 & 6.2 & -3.3 & 3.6 & -2.6 & 10.1 & 2.7 & -0.9 & 3.7 & 2.8 & 0.0 & 1.7 & 4.6 & -4.4 & 3.4 & 6.3 & 1.9\\
D2V & 0.0 & 8.0 & 0.0 & -8.2 & -2.3 & -25.3 & 6.1 & 28.1 & 0.7 & 2.1 & 8.3 & -4.5 & 4.1 & -3.0 & 2.6 & 1.6 & -0.6 & 3.3 & 1.8 & 0.0 & 1.3 & 13.9 & -5.1 & -3.7 & 6.9 & 1.3\\
APC & 3.1 & 6.3 & 0.0 & -8.4 & 0.0 & -25.9 & 4.1 & 29.4 & 1.1 & -0.7 & 7.5 & -4.6 & 4.6 & -2.6 & 9.7 & 2.3 & -1.0 & 4.7 & 2.8 & -0.7 & 0.2 & 5.4 & -4.5 & 1.0 & 6.9 & 2.0\\
mean & 0.7 & 5.9 & 0.0 & -4.5 & 0.4 & -25.4 & 7.1 & 29.2 & 1.7 & 0.9 & 8.3 & -4.0 & 3.4 & -2.9 & 9.1 & 2.5 & -0.9 & 4.8 & 2.8 & -0.7 & 0.2 & 5.4 & -4.5 & 1.2 & 7.0 & 1.7\\ 
\bottomrule
\end{tabular}
\fontsize{7}{9}\selectfont
\begin{tabular}{l||ccccc|cccccccccccc|ccccccccc}
\toprule
& \multicolumn{5}{c}{MSP-IMPROV} & \multicolumn{12}{|c|}{BIIC-NNIME} & \multicolumn{9}{c}{MSP-PODCAST}\\
\midrule
 & Ang & Sad & Neu & Hap & \textbf{Mac} & Ang & Fru & Disa & Sad & Fear & Neu & Sur & Exc & Hap & Rel & Joy & \textbf{Mac} & Ang & Sad & Disg & Cont & Fear & Neu & Sur & Hap & \textbf{Mac}\\
\midrule
XLS-R & 6.7 & 9.2 & -5.3 & 6.0 & 4.1 & -6.1 & -9.1 & -6.8 & 4.9 & 2.5 & 0.3 & 5.7 & 11.1 & 1.2 & -1.1 & 6.0 & 0.8 & 10.1 & -0.5 & -4.1 & -6.5 & 6.3 & -3.2 & 4.5 & 2.1 & 1.1\\
WavLM & 1.7 & 8.4 & -6.0 & 2.6 & 1.7 & -5.9 & -9.9 & -2.0 & 3.9 & 0.2 & 2.0 & 7.8 & 2.5 & 5.1 & 11.7 & 14.0 & 2.7 & 9.5 & -0.5 & -3.3 & -0.6 & 3.6 & -3.0 & 0.6 & 2.0 & 1.0\\
W2V2 R & 1.5 & 9.9 & -3.7 & 4.4 & 3.0 & -10.2 & -0.1 & -2.9 & -1.6 & 1.0 & 1.5 & 18.2 & 10.1 & -2.3 & -1.2 & 3.4 & 1.5 & 10.3 & -0.9 & -7.7 & -7.3 & 3.2 & -3.1 & 3.6 & 1.8 & 0.0\\
W2V2 & 1.3 & 8.6 & -5.8 & 1.6 & 1.4 & -8.7 & -4.7 & -8.0 & -0.3 & -0.2 & 0.2 & 4.2 & 4.4 & -2.8 & -1.6 & 6.1 & -1.0& 8.6 & -1.6 & -6.9 & -9.2 & 0.3 & -3.4 & 3.0 & 2.2 & -0.9\\
VQ-APC & -3.5 & 9.5 & -5.1 & -1.2 & -0.1 & -5.3 & -7.6 & -7.9 & 1.6 & -3.8 & -0.6 & 11.5 & 6.4 & -2.2 & 0.1 & 10.1 & 0.2 & 7.6 & -3.8 & 0.0 & -11.9 & 0.3 & -3.5 & 3.2 & 0.9 & -0.9\\
HuBERT & 4.4 & 7.1 & -6.3 & 2.7 & 1.9 & -5.7 & -1.9 & -5.6 & 1.5 & 0.0 & 1.3 & 4.7 & 9.7 & 2.3 & 3.2 & 7.9 & 1.6 & 9.4 & -1.3 & -5.7 & -4.9 & 1.8 & -3.1 & 2.3 & 1.8 & 0.0\\
DeCoAR 2 & 2.1 & 9.9 & -6.0 & 0.3 & 1.6 & -5.6 & -7.7 & -0.3 & 2.7 & -0.8 & 0.1 & 12.1 & 5.7 & 5.0 & -3.2 & 5.5 & 1.2 & 8.5 & -2.8 & -5.7 & -11.3 & 1.0 & -3.3 & 1.8 & 1.8 & -1.2\\
D2V & 2.0 & 10.0 & -5.5 & 1.9 & 2.1 & -6.5 & -3.9 & -5.5 & 4.9 & 2.3 & 0.0 & 5.7 & 5.6 & 0.8 & -1.0 & 1.4 & 0.3 & 9.1 & -1.2 & -4.9 & -6.1 & 1.4 & -3.0 & 5.1 & 1.8 & 0.3\\
APC & -0.4 & 10.6 & -6.3 & -1.2 & 0.7 & -5.1 & -7.3 & -4.3 & 2.5 & -1.4 & -0.3 & 9.7 & 6.7 & -2.2 & -3.3 & 8.9 & 0.3 & 8.1 & -3.6 & -1.2 & -11.8 & 0.8 & -3.6 & 3.7 & 0.9 & -0.8\\
mean & 0.6 & 9.3 & -6.1 & 1.6 & 1.3 & -6.9 & -5.7 & -3.6 & 2.0 & 0.7 & 0.2 & 9.6 & 6.5 & 1.3 & -0.2 & 6.7 & 1.0 & 8.0 & -2.7 & -2.6 & -8.7 & 1.3 & -3.4 & 3.3 & 1.4 & -0.4 \\
\bottomrule
\end{tabular}
\label{tab:emotion_f1_diff}
\vspace{-0.3cm}
\end{table*}

\subsubsection{Training data bias on emotion distribution}
We evaluate the bias in downstream training data distribution $\mathbf{d_d}$ by examining the difference in mean emotion distribution in training data (soft-label) for males and females. Take an emotion dataset with 4 emotions (neutral (N), anger (A), sadness (S), and happiness (H)) for example. We compute average training data distribution, $(N, A, S, H) = (0.2, 0.3, 0.4, 0.1)$ for females and $(N, A, S, H) = (0.1, 0.2, 0.4, 0.3)$ for males. Their training data bias is then $\mathbf{d_d} = (0.1, 0.1, 0.0, -0.2)$. For datasets with cross-fold validation, we compute the emotion distribution difference of the whole dataset.


\subsection{Datasets}
To make our evaluation comprehensive, we evaluate SER performance parity using six emotion datasets with diverse languages, speaker sources, and emotion types. 
The datasets include real-world datasets MSP-PODCAST (POD) \cite{Lotfian_2019_3}, BIIC-PODCAST (BPO) \cite{10388175}, and actor performance datasets MSP-IMPROV (IMP) \cite{Busso_2017}, IEMOCAP (IEM) \cite{Busso_2008_5}, BIIC-NNIME (NNI) \cite{Chou_2017}, CREMA-D (CRE) \cite{Cao_2014}. BIIC-NNIME and BIIC-Podcast are in Mandarin, and other datasets are in English.

We follow SpEAT, using the Morgan Emotional Speech Set (MESS) \cite{morgan2019categorical} for valence stimuli in upstream bias evaluation. We use the emotion datasets with both valence and speaker ID annotation in SER training (NNI, IMP, POD) as more valence stimuli for fair comparison. We use the Speech Accent Archive \cite{weinberger2011speech} for male and female stimuli. The selection criteria for stimuli are the same as those for SpEAT, with the goal of ensuring that differences in association with positive and negative valence do not stem from variations among the speakers.

\section{Result and Discussion}
\subsection{Downstream performance difference}
Table~\ref{tab:emotion_f1_diff} shows the difference in $F_1$ score between females and males on each dataset and each emotion. Due to space limitations, we report the result for models across 3 different training objectives and achieve top-9 performance in EMO-SUPERB. 
It shows that despite most ER models only exhibiting a slightly high macro-$F_1$ score for females, a high parity exists between the $F_1$ score for each emotion. 
For instance, in the BPO dataset, males exhibit approximately a 25\% higher $F_1$ score than females for neutral emotion, whereas they demonstrate a 29\% lower $F_1$ score for happy emotion. 

The bias observed in emotion $F_1$ scores varies across datasets but shows a consistent trend across emotion recognition models.
Specifically, we take BPO and POD as examples since they have identical emotion categories. 
While the POD dataset exhibits a slight SER bias $d_e$ in the happy and neutral emotion, the BPO dataset displays a significant $F_1$ score disparity in these categories, indicating substantial dissimilarity. 
Furthermore, all SER models trained with BPO consistently show the largest $d_e$ on happy and the smallest $d_e$ on neutral. Conversely, all SER models trained with the POD dataset consistently exhibit the highest $d_e$ values for emotion angry and the lowest for contempt. 
These observations underscore the potential correlation between gender bias in SER and the characteristics of the emotion dataset, with less influence from the upstream model. 
We conduct further analysis in section~\ref{sec:data_corr} and ~\ref{sec:upstream_corr}.

\begin{table}[t]
\centering
\fontsize{7}{9}\selectfont
\caption{(a) Corpus level (b) Valence gender bias on SER models in \%. Bold texts represent the model is the most biased within the same corpus.}
\vspace{-0.3cm}
\label{tab:d_c}
\setlength{\tabcolsep}{0.85pt}
\renewcommand{\arraystretch}{0.85}
\begin{tabular}{l|ccccccc|cccccc}
\toprule
 & \multicolumn{7}{c}{(a) Corpus level} & \multicolumn{6}{|c}{(b) Valence}\\
 & BPO & CRE & IEM & IMP & NNI & POD & avg & BPO & NNI & IEM & IMP & CRE & POD\\
\midrule
XLS-R & 9.9 & 5.8 & \textbf{4.2} & \textbf{6.8} & 5.0 & \textbf{4.7} & 6.1 & 15.1 & 25.4 & -4.3 & -9.9 & -1.4 & -0.5\\
WavLM & 9.9 & 5.1 & 3.9 & 4.7 & 5.9 & 2.9 & 5.4 & 10.7 & 37.2 & 1.2 & -7.5 & -10.5 & -5.2 \\
W2V & 10.3 & 5.8 & 3.1 & 4.9 & 4.1 & 4.6 & 5.5 & 26.4 & 11.4 & 2.2 & -11.9 & -4.8 & 7.0 \\
W2V2 R & 10.1 & 6.1 & 3.2 & 4.9 & 4.8 & \textbf{4.7} & 5.6 & 20.8 & 8.1 & 0.1 & -7.0 & -1.4 & 3.0\\
W2V2 & 10.9 & 3.5 & 2.7 & 4.3 & 3.7 & 4.4 & 4.9 & 27.5 & 24.8 & 9.4 & -8.4 & -6.7 & 9.9 \\
VQW2V & 8.9 & 4.7 & 3.0 & 5.3 & 3.6 & 4.1 & 4.9 & 26.4 & 17.2 & -1.6 & -8.6 & -6.3 & 1.2 \\
VQ-APC & 8.9 & 4.7 & 3.4 & 4.8 & 5.2 & 3.9 & 5.2 & 32.9 & 31.8 & -4.8 & -7.2 & 10.2 & 8.3 \\
TERA  & 8.4 & 5.6 & 3.3 & 5.0 & 5.1 & 4.6 & 5.3 & 33.6 & 34.9 & 4.6 & -9.5 & 1.8 & 9.2 \\
NPC & 8.9 & \textbf{6.8} & 2.7 & 5.6 & 5.4 & \textbf{4.7} & 5.7 & 44.2 & 22.9 & 3.8 & -8.9 & 11.1 & 19.2 \\
M CPC & \textbf{11.3} & 6.0 & 3.7 & 6.3 & \textbf{6.3} & 3.9 & \textbf{6.3} & 41.0 & 8.7 & -4.2 & -5.3 & 11.1 & 8.6 \\
Mock & 9.2 & 6.3 & 3.5 & 4.5 & 5.0 & 3.4 & 5.3 & 30.8 & 33.4 & 2.6 & -9.6 & 10.5 & 9.3 \\
HuBERT & 10.7 & 4.0 & 4.0 & 5.1 & 4.0 & 3.8 & 5.2 & 14.5 & 31.4 & 2.8 & -8.8 & -2.6 & 3.4 \\
FBANK & 7.6 & 0.6 & 3.2 & 2.6 & 1.3 & 0.7 & 2.7 & 36.0 & 5.4 & -17.8 & 3.5 & 0.0 & -1.3 \\
DeCoAR 2 & 8.8 & 4.6 & 3.1 & 4.6 & 4.4 & 4.5 & 5.0 & 34.3 & 21.4 & 1.8 & -11.8 & 1.7 & 12.5 \\
D2V & 9.8 & 4.1 & 2.9 & 4.9 & 3.4 & 4.1 & 4.9 & 28.9 & 13.0 & 0.8 & -10.1 & -7.4 & 2.4 \\
APC & 9.6 & 4.9 & 4.1 & 4.6 & 4.7 & 4.2 & 5.4 & 28.8 & 14.0 & -2.4 & -11.5 & 2.8 & 8.1 \\
\bottomrule
\end{tabular}
\vspace{-0.3cm}
\end{table}


We further compare the average $F_1$ score parity $d_c$ across models and datasets in Table~\ref{tab:d_c}(a). It shows that gender bias is closely related to datasets. SER models trained with BPO exhibit higher $d_c$ than those trained with other datasets, while SER models trained with IEM are less biased. Furthermore, the mean $d_c$ across datasets shows Modified CPC is more biased than all the other models, while XLS-R shows the most biased result on a majority of emotion datasets. Conversely, the baseline feature, FBank, shows less bias across corpora. This is caused by its low classification accuracy and recall of models trained with this feature. 

Also, we analyze the gender bias on valence $d_v$ by observing whether females have higher $F_1$ scores on positive valence and males have higher $F_1$ scores on negative valence. The result in Table~\ref{tab:d_c} (b) indicates that SER models trained with Chinese datasets (BPO and NNI) have higher $d_v$ than SER models trained with English datasets, and these models associate female speech with positive valence and male speech with negative valence. In contrast, SER models trained with English datasets demonstrate positive and negative $d_v$ values across different datasets, suggesting a more varied impact. 

Since the downstream model architecture is the same across all SER models trained with SSL fashion, the bias may only come from two possible sources: training data and upstream representation. We try to find out the most influential factor of gender bias via the following experiments.

\begin{table}[t]
\centering
\fontsize{7}{9}\selectfont
\caption{Pearson correlation coefficient between $F_1$-score gap on gender $d_e$ and training data distribution difference.}
\vspace{-1em}
\label{tab:data_corr}
\setlength{\tabcolsep}{3pt}
\renewcommand{\arraystretch}{0.85}
\begin{tabular}{l|cccccc}
\toprule
Model & BPO & CRE & IEM & IMP & NNI & POD \\
\midrule
XLS-R & 0.31 & 0.78 & 0.64 & 1.00 & 0.62 & 0.48 \\
WavLM & 0.43 & 0.88 & 0.58 & 0.92 & 0.66 & 0.40 \\
W2V & 0.40 & 0.74 & 0.73 & 0.70 & 0.39 & 0.34 \\
W2V2 R & 0.34 & 0.77 & 0.74 & 0.86 & 0.35 & 0.43 \\
W2V2 & 0.53 & 0.96 & 0.79 & 0.89 & 0.62 & 0.46 \\
VQW2V & 0.34 & 0.76 & 0.75 & 0.86 & 0.73 & 0.22  \\
VQ-APC & 0.47 & 0.69 & 0.50 & 0.61 & 0.61 & 0.38 \\
TERA & 0.44 & 0.76 & 0.66 & 0.80 & 0.60 & 0.40 \\
NPC & 0.52 & 0.69 & 0.63 & 0.65 & 0.46 & 0.46 \\
M CPC & 0.62 & 0.62 & 0.78 & 0.59 & 0.53 & 0.28 \\
Mock & 0.48 & 0.68 & 0.64 & 0.94 & 0.65 & 0.41 \\
HuBERT & 0.39 & 0.82 & 0.63 & 0.98 & 0.55 & 0.42\\
FBANK & 0.47 & 0.87 & 0.94 & 0.89 & 0.56 & 0.48 \\
DeCoAR 2 & 0.36 & 0.78 & 0.72 & 0.85 & 0.64 & 0.45\\
D2V & 0.48 & 0.76 & 0.70 & 0.87 & 0.63 & 0.40 \\
APC & 0.40 & 0.78 & 0.41 & 0.75 & 0.65 & 0.39 \\
\bottomrule
\end{tabular}
\vspace{-0.3cm}
\end{table}

\subsection{Downstream training data distribution}
\label{sec:data_corr}
We evaluate the Pearson correlation coefficient between \textit{training data distribution bias} $\mathbf{d_d}$ and \textit{the $F_1$-score gap between gender} $[d_{e_1}, d_{e_2}, ... d_{e_i}]$ on dataset with $i$ emotions $e_1, e_2 ... e_i$. 
Table~\ref{tab:data_corr} (a) reveals significant correlation coefficients for datasets featuring acted performances (CRE, IEM, IMP, NNI), suggesting a strong influence of data source on bias. 
Conversely, real-world datasets (BPO and POD) exhibit moderate correlation. 
This observation underscores the importance of considering the variance in emotion distribution across diverse social groups, as it might propagate to SER performance.

\begin{table}[b]
\centering
\fontsize{7}{9}\selectfont
\caption{Upstream representation bias $d_s$ measured on models with different representation aggregation methods. 'mean' denotes taking the average embedding over all layers. Other columns named by datasets denote using the weighted sum of embedding from ER models trained with the dataset.}
\vspace{-1em}
\label{tab:upstream_bias}
\setlength{\tabcolsep}{2.5pt}
\renewcommand{\arraystretch}{0.85}
\begin{tabular}{l|ccccccc}
\toprule
Model & mean & BPO & CRE & IEM & IMP & NNI & POD \\
\midrule
XLS-R & 1.06 & 1.08 & 1.06 & 1.06 & 1.06 & 1.06 & 1.08 \\
WavLM & 1.43 & 1.43 & 1.43 & 1.43 & 1.43 & 1.43 & 1.48 \\
W2V & 1.33 & 1.35 & 1.34 & 1.36 & 1.36 & 1.35 & 1.35 \\
W2V2 R & 0.29 & 0.79 & 0.86 & 0.89 & 0.87 & 0.76 & 0.83 \\
W2V2 & 0.58 & 0.58 & 0.56 & 0.58 & 0.57 & 0.58 & 0.57 \\
VQW2V & 0.67 & 0.65 & 0.61 & 0.64 & 0.65 & 0.65 & 0.66 \\
VQ-APC & 1.77 & 1.76 & 1.75 & 1.75 & 1.75 & 1.76 & 1.77 \\
TERA & 1.38 & 1.43 & 1.43 & 1.42 & 1.43 & 1.42 & 1.43 \\
NPC & 1.69 & 1.70 & 1.69 & 1.69 & 1.69 & 1.70 & 1.70 \\
M CPC & 0.53 & 0.55 & 0.57 & 0.52 & 0.52 & 0.52 & 0.59 \\
Mock & 1.05 & 1.06 & 1.06 & 1.06 & 1.06 & 1.07 & 1.06 \\
HuBERT & 0.99 & 1.03 & 0.99 & 0.99 & 0.99 & 0.99 & 1.02 \\
DeCoAR 2 & 1.49 & 1.46 & 1.48 & 1.48 & 1.48 & 1.48 & 1.46 \\
D2V & 0.45 & 0.58 & 0.49 & 0.44 & 0.45 & 0.46 & 0.50 \\
APC & 1.66 & 1.68 & 1.69 & 1.68 & 1.68 & 1.68 & 1.69 \\
\bottomrule
\end{tabular}
\end{table}

\begin{table}[t]
\centering
\fontsize{7}{9}\selectfont
\caption{Pearson correlation coefficient between downstream $F_1$ score gap on valence $d_v$ and upstream representation bias $d_s$ using different valence stimuli and aggregation methods (Aggr.).}
\vspace{-1em}
\setlength{\tabcolsep}{3pt}
\renewcommand{\arraystretch}{0.8}
\begin{tabular}{c|c|cccccc}
\toprule
Stimuli & Aggr. & BPO & CRE & IEM & IMP & NNI & POD \\
\midrule
\multirow{2}{*}{MESS}& Mean & 0.06 & 0.31 & 0.31 & -0.56 & 0.56 & 0.44 \\ 
& Weighted & 0.01 & 0.33 & 0.34 & -0.58 & 0.53 & 0.42 \\ 
\midrule
\multirow{2}{*}{NNIME}& Mean & -0.45 & -0.37 & -0.01 & 0.02 & -0.22 & -0.43 \\
& Weighted & -0.47 & -0.39 & -0.03 & 0.03 & -0.22 & -0.46\\ 
\midrule
\multirow{2}{*}{IMPROV} & Mean & 0.20 & 0.44 & 0.06 & -0.31 & 0.11 & 0.38\\ 
& Weighted & 0.13 & 0.45 & 0.07 & -0.33 & 0.09 & 0.32 \\
\midrule
\multirow{2}{*}{PODCAST} & Mean & -0.30 & -0.16 & -0.09 & 0.04 & -0.03 & -0.32 \\ 
& Weighted & -0.25 & -0.13 & -0.10 & 0.03 & 0.06 & -0.29 \\
\bottomrule
\end{tabular}
\label{tab:upstream_corr}
\vspace{-0.3cm}
\end{table}

\subsection{Upstream representation}
\label{sec:upstream_corr}
We first evaluate the upstream representation bias using SpEAT. As showcased in Table~\ref{tab:upstream_bias}, our findings unveil substantial levels of representation bias across most models, especially VQ-APC, NPC, and APC, which have larger $d_s$ values. In contrast, D2V and M CPC have smaller gender biases among these models. Moreover, our analysis reveals that aggregating layerwise representations via weighted sum yields consistent outcomes akin to mean pooling, underscoring the robustness and stability of our bias evaluation metric across diverse SER tasks.

We further evaluate the Pearson correlation coefficient between the downstream $F_1$ score gap on valence $d_v$ and upstream bias on valence $d_s$ among all models. Table~\ref{tab:upstream_corr} shows low or no correlation between upstream and downstream valence bias on different stimuli and aggregation methods. 
This implies gender bias in upstream representation might hardly propagate to downstream emotion classification tasks, which contradicts the conclusion in SpEAT. 
Two possible reasons might contribute to the difference between our work and SpEAT: 
(1) SpEAT trains the downstream valence prediction model with only 1800 speech samples, while we train multilabel emotion classification models and then calculate the difference between gender $F_1$ parity of positive and negative valence emotions, with at least 4000 speech samples per model. 
(2) SpEAT asserts that the group identified as positive valence by the pre-trained model tends to exhibit a similar association with positive valence in the downstream SER model, framing the challenge as a binary classification problem.
However, our analysis transcends binary classification by discussing the correlation between metrics representing continuous scales that reflect the extent of association between upstream and downstream bias. Our analysis suggests that we should use upstream bias metrics carefully as the bias might not reflect in application-level performance.

\section{Discussion and Limitation}
Gender is a spectrum rather than a solely male/woman binary \cite{bass2018rethinking}.
However, existing speech emotion classification dataset datasets only have labels on binary gender. 
Including a broader range of gender identities in the dataset would better reflect the reality of human diversity and improve the performance and fairness of the models, which can be our future work.

\section{Conclusion and Future Work}
This work provides extensive insights into the gender bias in SER models trained with the SSL paradigm. We identify that females exhibit superior overall SER performance compared to males. Also, a substantial gender-wise performance gap exists across individual emotions. Furthermore, our investigation underscores the influence of dataset characteristics, revealing Mandarin datasets to exhibit a pronounced bias toward valence compared to their English counterparts. Importantly, we find that downstream training data distribution plays a pivotal role in exacerbating gender bias, while upstream representation exerts minimal influence. 
These findings have far-reaching implications for developing ER technologies. Our forthcoming efforts aim to rectify biases inherent in these ER systems. Our research sets the stage for more inclusive and ethical approaches to designing and deploying AI systems.

\bibliographystyle{IEEEtran}
\bibliography{mybib}

\end{document}